\documentclass[12pt]{article}
\usepackage{graphicx}

\newcommand\pubnumber{FERMILAB-CONF-11-028-PPD}
\newcommand\pubdate{\today}

\newcommand {\asld} {\ensuremath{a^d_{\mathrm{sl}}}}
\newcommand {\asls} {\ensuremath{a^s_{\mathrm{sl}}}}
\newcommand {\aslq} {\ensuremath{a^q_{\mathrm{sl}}}}
\newcommand {\aslb} {\ensuremath{A^b_{\mathrm{sl}}}}
\newcommand {\Bd} {\ensuremath{B^0_d}}
\newcommand {\Bs} {\ensuremath{B^0_s}}

\def\lancs{Physics Department, Lancaster University, UK \\
On behalf of D\O\ Collaboration}

\def\Title#1{\begin{center} {\Large #1 } \end{center}}
\def\Author#1{\begin{center}{ \sc #1} \end{center}}
\def\Address#1{\begin{center}{ \it #1} \end{center}}

\newcommand\pubblock{\rightline{\begin{tabular}{l} \pubnumber\\
         \pubdate  \end{tabular}}}
\newenvironment{Abstract}{\begin{quotation}  }{\end{quotation}}
\newenvironment{Presented}{\begin{quotation}
      \begin{center}\begin{large}}{\end{large}\end{center} \end{quotation}}

\def\beq{\begin{equation}}
\def\eeq#1{\label{#1}\end{equation}}
\def\eeqn{\end{equation}}

\def\beqa{\begin{eqnarray}}
\def\eeqa#1{\label{#1}\end{eqnarray}}
\def\eeqan{\end{eqnarray}}

\let\bar=\overbar

\def\O{{\cal O}}

\def\Dslash{\not{\hbox{\kern-4pt $D$}}}
\def\dslash{\not{\hbox{\kern-2pt $\del$}}}

\def\msb{{\bar{\ssstyle M \kern -1pt S}}}

\begin{document}
\begin{titlepage}
\pubblock

\vfill
\Title{Measurement of $\phi_s$ at D{\O}\ Experiment}
\vfill
\Author{ Guennadi Borissov}
\Address{\lancs}
\vfill
\begin{Abstract}
Recent measurements of the D\O\ experiment related to the search for new phenomena beyond the Standard Model are reviewed. The new measurement of the like-sign dimuon charge asymmetry reveals a 3.2$\sigma$ deviation from the SM prediction, while the updated study of the $B_s \to J/\psi \phi$ decay demonstrates a better agreement with the SM. All experimental results on the $CP$ violation in mixing are currently consistent with each other. The D\O\ collaboration has much more statistics to analyze, and all these results can be significantly improved in the future.
\end{Abstract}
\vfill
\begin{Presented}
Proceedings of CKM2010, the 6th International Workshop on the CKM  Unitarity Triangle, University of Warwick, UK, 6-10 September 2010
\end{Presented}
\vfill
\end{titlepage}
\def\thefootnote{\fnsymbol{footnote}}
\setcounter{footnote}{0}
%

The time evolution of the neutral $B_q - \bar{B}_q$ meson system $(q = d,s)$ is described by the complex mass matrix:
\begin{equation}
\| M \| =
\left[
\begin{array} {cc}
M_q & M^{12}_q \\
(M_q^{12})^* & M_q
\end{array}
\right]
- \frac{i}{2}
\left[
\begin{array}{cc}
\Gamma_q & \Gamma_q^{12} \\
(\Gamma_{q}^{12})^* & \Gamma_q
\end{array}
\right]
\end{equation}
In addition to the mass $M_q$ and width $\Gamma_q$ of $B_q$ meson, the following observables can be defined for this system:
\begin{eqnarray}
\Delta M_q & = & M_H - M_L \approx 2 |M_q^{12}| \nonumber \\
\Delta \Gamma_q & = & \Gamma_L - \Gamma_H \approx 2 | \Gamma_q^{12}| cos(\phi_q) \nonumber \\
\phi_q & = & \arg (- M_q^{12}/\Gamma_q^{12})
\end{eqnarray}
A violation of the $CP$ symmetry is caused by a non-zero value of the phase $\phi_q$. The Standard Model predicts very small values of $\phi_d$ and $\phi_s$ which are significantly below current experimental sensitivity \cite{lenz}. The new physics can significantly change them \cite{asl-theo}. Therefore, a deviation from zero in these phases would be an unambiguous indication of the contribution of new physics.

The phase $\phi_q$ can be extracted from the charge asymmetry of semileptonic $B_q$ decays, from the like-sign dimuon charge asymmetry and from the decay $B_s \to J/\psi \phi$. The D\O\ experiment performs all these measurements and its results on the $CP$ violating phase $\phi$ are presented in this paper.

The charge asymmetry \aslq~ for ``wrong-charge"
semileptonic $B^0_q$-meson decay induced by oscillations is defined as
\begin{equation}
\aslq = \frac{\Gamma(\bar{B}^0_q(t)\rightarrow \mu^+ X) -
              \Gamma(    {B}^0_q(t)\rightarrow \mu^- X)}
             {\Gamma(\bar{B}^0_q(t)\rightarrow \mu^+ X) +
              \Gamma(    {B}^0_q(t)\rightarrow \mu^- X)}.
              \label{aslq}
\end{equation}
This quantity is independent of the decay time $t$, and can be expressed as
\begin{equation}
\aslq = \frac{\left| \Gamma_q^{12} \right|}{\left| M_q^{12} \right|} \sin \phi_q =
\frac{\Delta \Gamma_q}{\Delta M_q} \tan \phi_q.
\label{phiq}
\end{equation}

The value of \asld~ measured at B-factories is
$\asld = -0.0047 \pm 0.0046$~\cite{hfag}. The value of \asls~ is measured uniquely by the D\O\ experiment \cite{d0-asls}. The semileptonic decays $B_s \to \mu \nu D_s$ with $D_s \to \phi \pi$ or $D_s \to K^* K$ are selected. Using the integrated luminosity of 4 fb$^-1$ the D\O\ experiment reconstructed  about 81.4K $D_s \to \phi \pi$ and 33.6K $D_s \to K^* K$ decays. The obtained result is consistent with the SM prediction:
\begin{equation}
\asls = -0.0017 \pm 0.0091 \mbox{~(stat)} ^{+0.0014}_{-0.0015} \mbox{~(syst)}
\end{equation}

The like-sign dimuon charge asymmetry
$\aslb$ for semileptonic decays of $b$ hadrons produced in
proton-antiproton ($p \bar{p}$) collisions is defined as
\begin{equation}
\aslb \equiv \frac{N^{++}_{b} - N^{--}_{b}}{N^{++}_{b} + N^{--}_{b}},
\end{equation}
where $N^{++}_{b}$ and $N^{--}_{b}$ are the numbers of events
containing two $b$ hadrons that decay semileptonically, producing two positive or
two negative muons, respectively, with only the direct semileptonic decays $b \rightarrow \mu X$
considered in the definition of $N^{++}_{b}$ and $N^{--}_{b}$.
The asymmetry $\aslb$ can be expressed~\cite{Grossman} as
\begin{equation}
\aslb = \frac{f_d Z_d \asld + f_s Z_s \asls}{f_d Z_d + f_s Z_s},
\label{Ab1}
\end{equation}
where
\begin{eqnarray}
Z_q & \equiv & \frac{1}{1 - y_q^2} - \frac{1}{1 + x_q^2}, \\
y_q & \equiv & \frac{\Delta \Gamma_q}{2 \Gamma_q}, ~~ x_q \equiv \frac{\Delta M_q}{\Gamma_q}.
\end{eqnarray}
with $q = d, s$. The quantities $f_d$ and $f_s$ are the
production fractions for $\bar{b} \rightarrow \Bd$ and
$\bar{b} \rightarrow \Bs$ respectively. Using the current experimental values \cite{pdg} of the quantities entering in (\ref{Ab1}) we obtain
\begin{equation}
\label{aslb}
\aslb = (0.506 \pm 0.043) \asld + (0.494 \pm 0.043) \asls.
\end{equation}

The D\O\ collaboration reports a new measurement of \aslb~ using an integrated luminosity of 6.1 fb$^{-1}$ \cite{d0-asl}. In this analysis the events with two muons of the same charge and with one muon are selected, and the inclusive like-sign dimuon and single muon charge asymmetries are measured. In addition to the direct $b \to \mu X$ decay many other background processes contribute in these samples. Only the direct semileptonic decay $b \to \mu X$ can produce the $CP$ violating charge asymmetry; the charge asymmetry of all other processes is detector-related. All background contributions in this analysis are determined experimentally with the reduced input from simulation. The sample of events with single muon is used to constrain and control the background. With this approach the systematic uncertainties are considerably reduced.

A unique feature of the D\O\ experiment is a regular reversal of magnet polarities. Due to this reversal the charge asymmetry due to the muon reconstruction is suppressed and the related systematic uncertainty is considerably reduced. The remaining dominant source of background asymmetry comes from the difference in the interaction cross section of positive and negative kaons with the detector material. This asymmetry is measured experimentally.

\begin{figure}[tbh]
\begin{center}
\includegraphics[width=0.50\textwidth]{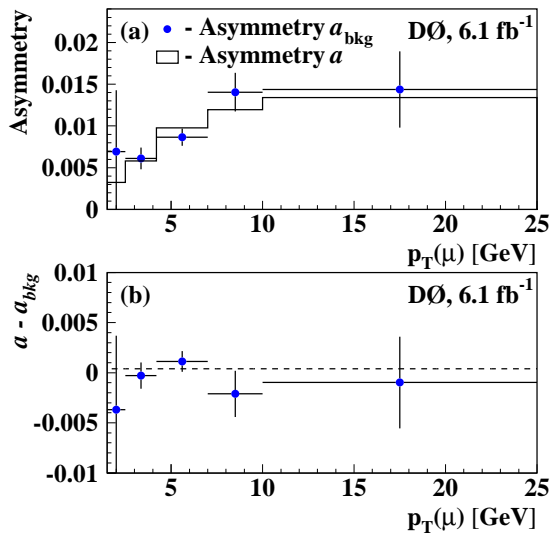}
\caption{(a) The background asymmetry (points with error bars) is compared to the measured asymmetry a of the inclusive muon sample (shown as
histogram, since the statistical uncertainties are negligible). The asymmetry from CP violation is negligible compared to the background in the inclusive muon sample; (b) the difference between the observed and expected asymmetries. The horizontal dashed line shows the mean
value of this difference.}
\label{fig3}
\end{center}
\end{figure}

The charge asymmetry of the inclusive muon sample is mainly determined by the background processes, and the possible contribution of the signal is suppressed. Therefore, the inclusive muon charge asymmetry is used to verify the measurement of the background asymmetries. Figure \ref{fig3} shows the comparison of the inclusive muon charge asymmetry and the asymmetry expected from the background processes as function of muon transverse momentum $p_T(\mu)$. A good agreement between the observed and expected asymmetries confirms the validity of the measurement method.

\begin{figure}[t]
\begin{center}
\includegraphics[width=0.50\textwidth]{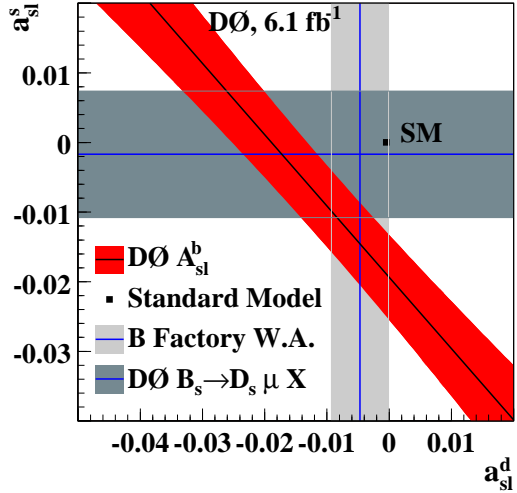}
\caption{Comparison of \aslb~ with the standard model prediction for \asld~ and and \asls. Also shown are the existing measurements of \asld~ \cite{hfag} and \asls~ \cite{d0-asls}. The error bands represent the $\pm 1$ standard deviation uncertainties on each individual measurement.}
\label{fig4}
\end{center}
\end{figure}

After subtracting the background asymmetries from the inclusive dimuon charge asymmetry, the resulting like-sign dimuon charge asymmetry is found to be
\begin{equation}
A_{sl}^b = -0.957 \pm 0.251 \mbox{~(stat)} \pm 0.146 \mbox{~(syst)}.
\end{equation}
This result differs from the SM prediction by $\sim 3.2 \sigma$. The asymmetry $A_{sl}^b$ can be expressed as a linear combination of the semileptonic charge asymmetries \asld~ and \asls, see (\ref{aslb}). Therefore, the obtained result is presented in Fig.~\ref{fig4} as a diagonal band in a two dimensional plane of \asls~ versus \asld. This result is in a good agreement with the direct measurements of the \asls~ and \asld~ asymmetries \cite{hfag, d0-asls}.

The decays $B_s \to J/\psi \phi$ offers a complementary possibility to probe the $CP$ violation in mixing of $B_s$ meson. The $CP$ violation in this decay is described by the phase $\phi^{J/\psi \phi}$. Within the Standard Model this phase is related with the angle $\beta_s$ of the $(bs)$ unitarity triangle and should be very small \cite{lenz}.
However, it can be significantly modified by the new physics contribution and its variation is the same as for the phase $\phi_s$ \cite{lenz}.

\begin{figure}[t]
\begin{center}
\includegraphics[width=0.50\textwidth]{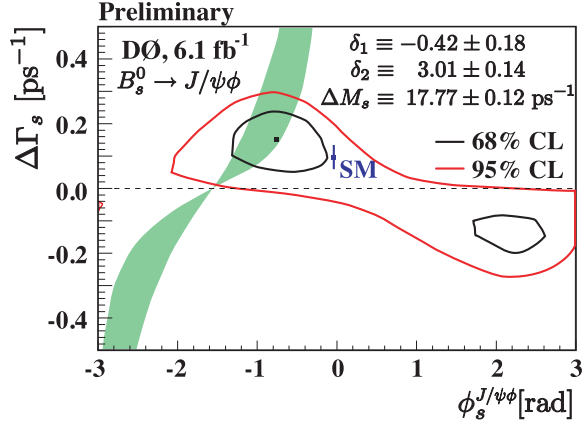}
\caption{68\% and 95\% C.L. contours in the plane $\Delta \Gamma_s - \phi_s$. Also shown is the 68\% contour
from the D0 dimuon charge asymmetry analysis \cite{d0-asl}. The comparison is made under the assumption of a single source of the CP violation in the $B_s \-- \bar{B}_s$ mixing.}
\label{fig6}
\end{center}
\end{figure}

The D\O\ collaboration presents an update of the analysis of $CP$ violation in $B_s \to J/\psi \phi$ decay using an integrated luminosity of 6.1 fb$^{-1}$ \cite{d0-psiphi}. We obtain
\begin{eqnarray}
\tau_s & = & 1.47 \pm 0.04 \mbox{~(stat)} \pm 0.01 \mbox{~(syst)~ ps}, \nonumber \\
\Delta \Gamma_s & = & 0.15 \pm 0.06 \mbox{~(stat)} \pm 0.01 \mbox{~(syst)~ ps}^{-1}.
\end{eqnarray}
The result on the $CP$ violating phase $\phi^{J/\psi \phi}$ is shown in Fig.~\ref{fig6}. The 68\% C.L. constraint on the phase $\phi_s$ derived from the like-sign dimuon charge asymmetry is also shown in Fig.~\ref{fig6}. It can be seen that both results are consistent, although the measurement of $\phi_s$ phase from $B_s \to J/\psi \phi$ is better consistent with the Standard Model.

In conclusion, the D\O\ experiment is very active the searches for new physics beyond the Standard Model. The new measurement of the like-sign dimuon charge asymmetry performed by the D\O\ experiment reveals a 3.2$\sigma$ deviation from the SM prediction, while the updated study of the $B_s \to J/\psi \phi$ decay demonstrates a better agreement with the SM. Nevertheless, all experimental results are currently consistent with each other and further study with improved precision is required to reveal a possible contribution of new physics. The uncertainty of all measurements included in this report is statistically dominated, and all results can be improved in the future with the increase of collected statistics.

\end{document}